\documentclass[sigconf]{acmart}
\usepackage{lipsum}
\usepackage{multicol}
\usepackage{graphicx}
\usepackage{fancyvrb}
\usepackage[skip=0pt]{subcaption}
\usepackage[skip=6pt]{caption}
\usepackage{threeparttable}
\usepackage{mathtools}
\usepackage{adjustbox}
\usepackage{float}
\usepackage{colortbl}
\usepackage{tabularx}
\usepackage{amsmath}
\usepackage{hyperref}
\usepackage{flushend}
\usepackage{booktabs, dcolumn}
\usepackage{pifont}
\usepackage{wasysym}
\usepackage{tablefootnote}
\usepackage{mdframed}
\usepackage{fontawesome}
\usepackage[normalem]{ulem}
\usepackage{balance}
\usepackage{soul}
\definecolor{searchxblue}{rgb}{0.192, 0.651, 0.835}
\definecolor{easy}{rgb}{0.99, 0.99, 0.59} 
\definecolor{difficult}{rgb}{1.0, 0.7, 0.0} 

\usepackage{tikz}

\newcolumntype{D}[1]{>{\centering\arraybackslash}p{#1}}
\newcolumntype{L}[1]{>{\arraybackslash}p{#1}}
\AtBeginDocument{%
  \providecommand\BibTeX{{%
    \normalfont B\kern-0.5em{\scshape i\kern-0.25em b}\kern-0.8em\TeX}}}

\VerbatimFootnotes

\title{Searching to Learn with Instructional Scaffolding}

\author{Arthur C\^{a}mara, Nirmal Roy, David Maxwell, Claudia Hauff}
\affiliation{
    \institution{Delft University of Technology}
    \city{Delft}
    \country{The Netherlands}
}
\email{{a.barbosacamara, n.roy, d.m.maxwell, c.hauff}@tudelft.nl}

\newcommand*\circled[1]{\tikz[baseline=(char.base)]{\node[shape=circle,fill=searchxblue,inner sep=1pt] (char) {\textcolor{white}{#1}};}}

\newcommand{\sal}{SAL}
\newcommand{\searchx}{\texttt{Search}\color{searchxblue}\texttt{X}\color{black}}
\newcommand{\scentbar}{\texttt{ScentBar}}
\newcommand{\qe}{\texttt{AQE$_{\texttt{SC}}$}}
\newcommand{\visual}{\texttt{CURATED$_{\texttt{SC}}$}}
\newcommand{\progress}{\texttt{FEEDBACK$_{\texttt{SC}}$}}
\newcommand{\control}{\texttt{CONTROL}}
\newcommand{\car}{TREC CAR}
\newcolumntype{R}[2]{%
    >{\adjustbox{angle=#1,lap=\width-(#2)}\bgroup}%
    l%
    <{\egroup}%
}

\setcopyright{rightsretained}
\copyrightyear{2021}
\acmYear{2021}

\acmConference[CHIIR '21]{Proceedings of the 2021 ACM SIGIR Conference on Human Information Interaction and Retrieval}{March 14--19, 2021}{Canberra, ACT, Australia}
\begin{document}

\fancyhead{}
\thanks{This research has been supported by \textit{NWO} projects \textit{SearchX} (639.022.722) and \textit{Aspasia} (015.013.027).}

\copyrightyear{2021} 
\acmYear{2021} 
\acmConference[CHIIR '21]{Proceedings of the 2021 ACM SIGIR Conference on Human Information Interaction and Retrieval}{March 14--19, 2021}{Canberra, ACT, Australia}
\acmBooktitle{Proceedings of the 2021 ACM SIGIR Conference on Human Information Interaction and Retrieval (CHIIR '21), March 14--19, 2021, Canberra, ACT, Australia}
\acmDOI{10.1145/3406522.3446012}
\acmISBN{978-1-4503-8055-3/21/03}

\begin{abstract}

Web search engines are today considered to be the primary tool to assist and empower \textit{learners} in finding information relevant to their learning goals---be it learning something new, improving their existing skills, or just fulfilling a curiosity. While several approaches for improving search engines for the learning scenario have been proposed (e.g. a specific ranking function), \textit{instructional scaffolding} (or simply \textit{scaffolding})---a traditional learning \emph{support} strategy---has not been studied in the context of search as learning, despite being shown to be effective for improving learning in both digital and traditional learning contexts. When scaffolding is employed, instructors provide learners with support throughout their autonomous learning process. 
We hypothesize that the usage of scaffolding techniques within a search system can be an effective way to help learners achieve their learning objectives \emph{whilst searching}. As such, this paper investigates the incorporation of \textit{scaffolding} into a search system employing three different strategies (as well as a control condition): \textit{(i)}~\qe{}, the automatic expansion of user queries with relevant subtopics; \textit{(ii)}~\visual{}, the presenting of a manually curated static list of relevant subtopics on the search engine result page; and \textit{(iii)}~\progress{}, which projects real-time feedback about a user's exploration of the topic space on top of the \visual{} visualization. 
To investigate the effectiveness of these approaches with respect to human learning, we conduct a user study ($N=126$) where participants were tasked with searching and learning about topics such as \emph{`genetically modified organisms'}. 
We find that \textit{(i)} the introduction of the proposed scaffolding methods in the proposed topics does not significantly improve learning gains. However, \textit{(ii)} it does significantly impact search behavior. Furthermore, \textit{(iii)} immediate feedback of the participants' learning (\progress{}) leads to undesirable user behavior, with participants seemingly focusing on the feedback gauges instead of learning.

\end{abstract}

\maketitle

\section{Introduction}\label{sec:Introduction}

Search and sensemaking are an intricate part of a user's learning process. For many learners today this is synonymous with accessing and ingesting information through web search engines~\cite{selwyn2008investigation,biddix2011convenience,nicholas2011google}. Despite this, web search engines are not equipped to support users in the type of complex searches often required in learning situations\footnote{As a concrete complex search example, in our experiments we ask participants to learn about \emph{radiocarbon dating considerations} (among other topics).}~\cite{golovchinsky2012future,hassan2014supporting,marchionini2006exploratory}. The process of what is now known as \emph{Search as Learning (\sal{})}~\cite{collinsthompson_et_al:DR:2017:7357} was first formally defined by~\citet{marchionini2006exploratory} as an iterative process, \emph{mediated by a search system}, where learners purposefully engage by reading, scanning and processing a large number of documents with the ultimate goal of gaining knowledge about one specific learning objective. The finding, understanding, analyzing and evaluation~\cite{Anderson2001,urgo2019anderson} of documents containing information relevant to answering this question is a time-consuming and cognitively demanding process. 

Recently, a number of different research efforts have been devoted to the area of \sal{}, such as: \textit{(i)} the influence of user characteristics and user strategies on learning while searching~\cite{o2020role,pardi2020role,roy2020chiir,gadiraju2018analyzing,lu2017personalized,liu2018information}; \textit{(ii)} the exploration of user behavior during learning-oriented search sessions~\cite{roy2020chiir,Eickhoff2014,moraes2018contrasting,gadiraju2018analyzing}; \textit{(iii)} the prediction/observation of how knowledge changes over time and across different cognitive levels of learning~\cite{roy2020chiir,liu2019investigation,kalyani2019understanding,wildemuth2004effects,yu2018predicting}; \textit{(iv)} the measuring of learning during searches~\cite{Eickhoff2014,bhattacharya2018relating,von2019metacognitive,bhattacharya2019measuring,wilson2013comparison}; and \textit{(v)} the design of retrieval algorithms for learning-oriented search tasks and user interface components~\cite{syed2017optimizing,syed2020improving,syed2018exploring}. Despite the large number of prior works in the \sal{} field, only a small number have so far explored the \emph{adaptation of the search system itself} to improve learning outcomes.

During a learning-oriented search session, realizing \textit{what they do not know} about a topic is a key hurdle for learners to overcome. Previous work~\cite{yu2018predicting} has shown that learners, on average, are aware of only 40\% of the different aspects pertaining to a topic before the search session commences. To counter this issue, the learning sciences provide us with the concept of \emph{instructional scaffolding} for a classroom environment~\cite{merrill1992effective, brown1986guided, rogoff1986adult, vygotsky1980mind}. Using scaffolding, an instructor or teacher provides \emph{guidance} to learners through various means in order for them to achieve their learning goals. During the early stages of learning, these scaffolds provide plenty of structure and direction. Over time however, the responsibility of identifying core concepts about a topic shifts from the scaffolding to the learner. By the end of the learning process, the scaffold is withdrawn as no more guidance should be required.
  
When translating the idea of instructional scaffolding to digital learning, \citet{Hill2001education} proposed a number of different scaffolding components. Of special interest to us are the so-called \textit{conceptual scaffolds} (analogous to \textit{topical outlines}), designed to ``assist the learner in deciding what to consider or to prioritize what is important.'' In this paper, we explore to what extent conceptual scaffolds---which have been shown to be beneficial for human learning in digital learning environments---are beneficial for learning while searching. 


To this end, we propose three different strategies of incorporating scaffolding into learners' search sessions:  \textit{(i)}~\qe{}, the \textit{automatic} \textit{expansion} of users' \textit{queries} with relevant subtopics (i.e. key aspects of the topic to learn more about) as predefined by an expert; \textit{(ii)}~\visual{}, the presentation of a manually curated static list of relevant subtopics on the search engine result page, as also discussed recently by~\citet{smith2019knowledge} (in contrast to \qe{} the learner here is explicitly aware of the subtopics related to the main topic); and \textit{(iii)}~\progress{}, which projects real-time feedback about the users' exploration of the topic space on top of the \visual{} visualization. This is inspired by recent works like \texttt{ScentBar}~\cite{umemoto2016scentbar} and \citet{von2019metacognitive}, who posit that a better calibration of one's self-assessment of learning during search sessions can be achieved through the provision of automatically generated feedback that indicates learning progress.

We implemented these scaffolding variants on top of the \searchx{} framework~\cite{putra2018searchx}, and conducted an inter-subject study, where $126$ participants were randomly assigned to one of four conditions (the three variants introduced above, plus \control{}, a standard search interface) to assess how conceptual scaffolds impact human learning while searching. By measuring the participants' knowledge before and after each learning-oriented search session, we were able to measure their \emph{knowledge gain}. With this \textit{Interactive Information Retrieval (IIR)} experiment, we aim to answer the following research questions:
 
\begin{description}
    \item[RQ1] Is conceptual scaffolding beneficial to improve learners' knowledge gain compared to a standard search system setup?
    \item[RQ2] When scaffolding is introduced, to what extent does learners' search behavior change?
\end{description}

Our main findings can be summarized as follows. \textit{(i)} The proposed scaffolding methods are shown to not be significantly effective for increasing learners' knowledge gain, with gains ranging from 30\% to a detrimental effect of 7\%, when compared to the control condition. \textit{(ii)} The type of scaffold has a significant impact on learners' search behavior. We also show the participants' queries to be heavily influenced by the scaffolding components. \textit{(iii)} Participants in the \visual{} and \progress{} conditions were more engaged with the platform, and issued more queries, viewed more documents, and spent more time searching. At the same time, the \progress{} cohort exhibited behavior indicating that they focused on the feedback gauge more than the actual learning process.

\section{Related Work}
We now discuss the main findings in the \sal{} literature, which has been inspired by the observation that learners increasingly turn to the web (and thus web search engines) to support their learning needs~\cite{rowlands2008google, nicholas2011google}.

\subsubsection*{Influence of user characteristics and task characteristics on learning.} \citet{o2020role} explored the impact of domain expertise (experts vs. non-experts) on learning, with expertise determined based on participants' self-reported frequency of searches for historical information. No significant differences in learning outcomes were found between the two groups. This is in contrast to \citet{gadiraju2018analyzing}, where slightly higher knowledge gains for participants with less prior knowledge were observed. \citet{roy2020chiir} investigated \emph{when} during a search session learning takes place, and did observe differences between expert and non-expert learners, specially towards the end of the search session. Other than domain expertise, learners' cognitive abilities (such as working memory capacity and reading comprehension ability~\cite{pardi2020role}) were found to be predictive of learning outcomes. In terms of user strategies for learning, \citet{liu2018information} found learners who adapt their source selection to the type of task at hand (encyclopedia-style documents for receptive tasks and Q\&A documents for critical tasks) to have better learning outcomes than learners who do not adapt their source selection strategy. \citet{kalyani2019understanding} explored to what extent a search task's cognitive learning level (based on the revised Bloom's taxonomy~\cite{anderson1993computer}) impacts user behavior: as a trend, the higher the cognitive level of the search task (such as \emph{remembering} vs. \emph{applying}), the larger the amount of search interactions. Based on this insight, the authors proposed to train models in a supervised manner to automatically determine the complexity of a user's information needs. This in turn could lead to adaptive search systems that are optimized for learning.

\subsubsection*{Proxy measures of learning} The vast majority of the aforementioned works measure knowledge gain through knowledge tests: those are often multiple-choice tests, but manually annotated user summaries~\cite{o2020role, wilson2013comparison,liu2018information} and mind maps~\cite{liu2019investigation} have been explored as well. Knowledge tests are admitted before and after the search session in order to determine the knowledge gains throughout the session. However, in order to build search systems that are adaptive to users' learning needs, we require \textit{scalable and easy to collect behavioral metrics} that are predictive of knowledge gain. Past works have explored which behaviors can be considered to be predictive of learning. Document dwell time was found to be indicative of learning~\cite{Eickhoff2014,CollinsThompson2016}, as well as the number of \textit{Search Engine Results Page (SERP)} clicks~\cite{CollinsThompson2016}, the occurrence of contextually relevant imagery~\cite{syed2018exploring}, and the diversity of the domains present among the top-ranked documents~\cite{Eickhoff2014}. \citet{pardi2020role} studied the relationship between the kind of documents users dwell on and their learning outcomes. They found \emph{text-dominated} documents to be more effective for learning than video documents (this though is in contrast to the findings by~\citet{moraes2018contrasting}, who find high-quality video material to yield higher learning outcomes than searching). In contrast to the aforementioned studies that considered just a handful of predictors, \citet{yu2018predicting} evaluated approximately 70 search-based features as predictors of learning; individually they were found to be only weakly correlated with knowledge gain, though some (document dwell time, query complexity) were more predictive than others. Although unarguably less scalable, a number of eye-tracking measures (such as the duration or reading fixations within documents) have also shown to be predictive of learning outcomes~\cite{bhattacharya2018relating,bhattacharya2019measuring}. Finally, we note that~\citet{von2019metacognitive} also found learners to be able to estimate their learning performance with increasing accuracy as the search session progresses.

\begin{figure*}[h]
  \includegraphics[width=0.85\textwidth]{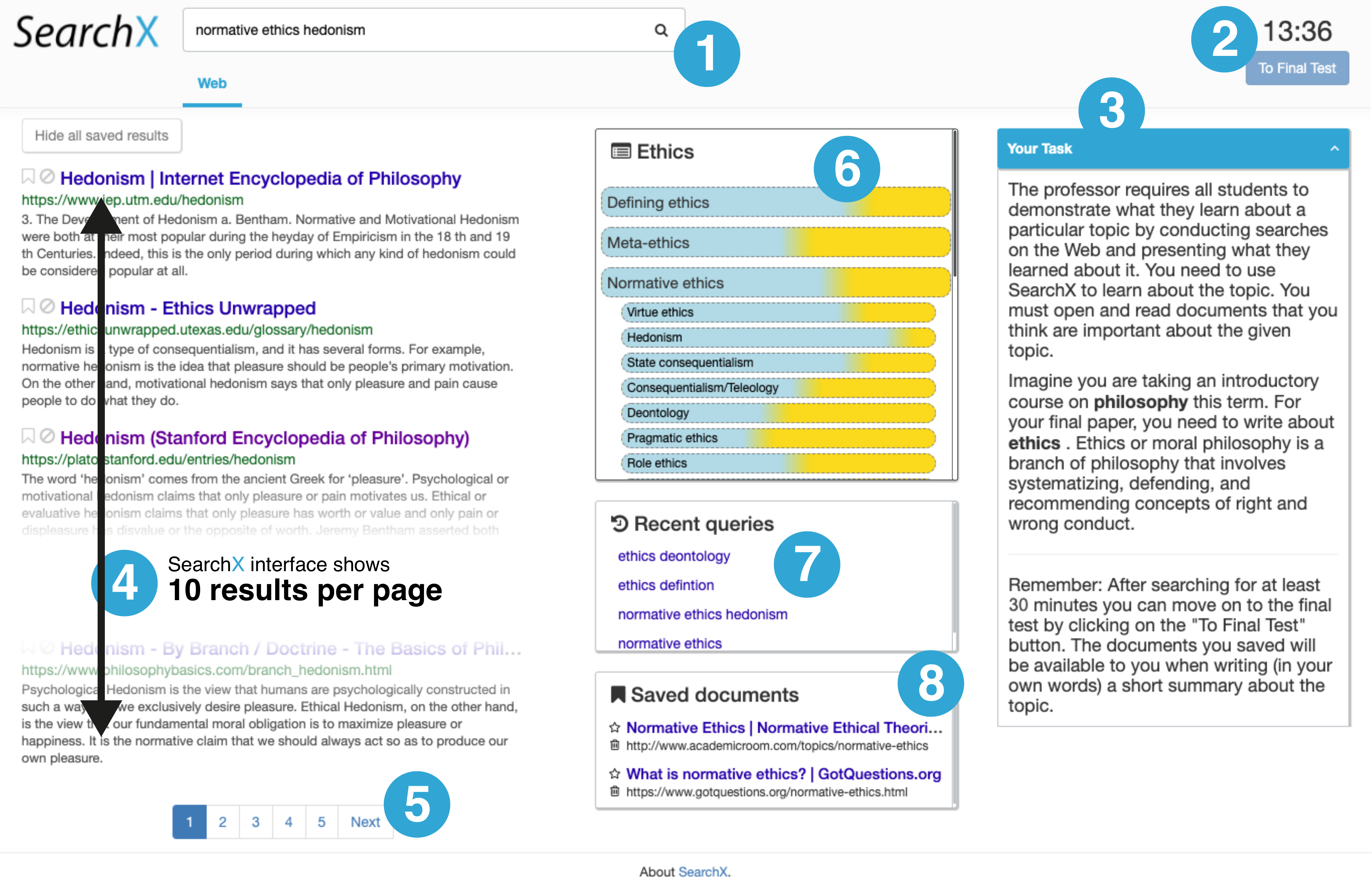}
  \caption{The \searchx{} interface: the eight annotated interface components are described in Section~\ref{sec:searchx_scaffolding}. Note that the scaffolding component (displaying the \texttt{Ethics} topic) shows the \progress{} scaffolding variant---complete with yellow progress gradients.}
  \label{fig:interface}
\end{figure*}

\subsubsection*{Retrieval system adaptation} Few works have explored so far the adaptation of the retrieval system itself to support learning. \citet{syed2017optimizing} designed a retrieval algorithm specifically for vocabulary learning by ranking documents according to their keyword density of the vocabulary items to learn. The user evaluation showed that at least for some topics, results with a higher keyword density leads to significantly higher learning gains (a follow-up study showed this to hold over a long period of time as well~\cite{syed2018exploring}). However, it should be noted that the user study fixed the documents to read for each topic, instead of allowing participants to search and adapting the retrieved results on the fly. Recently, \citet{syed2020improving} investigated whether automatic question generation can be utilized to improve learning outcomes \emph{whilst reading} a document. Although the improvement in learning outcome was limited to learners with low levels of prior knowledge, it is not far fetched to imagine such an interface component to also be incorporated in a search system.

\subsubsection*{Visualization of search progress} Lastly, we want to point to the work on \scentbar{} by~\citet{umemoto2016scentbar} which---though unrelated to \sal{}---inspired one of our scaffolding variants (\progress{}): it is a query suggestion interface that visualizes to what extent information relevant to the information need remains unexplored. A user study on a number of intrinsically diverse tasks showed that users were indeed better able to determine when to stop searching for relevant information when the amount of missed information was made visible to them. 
\section{Instructional Scaffolding in  \texttt{SearchX}}\label{sec:searchx_scaffolding}
We implemented our scaffolding variants as part of \searchx{}~\cite{putra2018searchx}, a modular, open-source search framework which provides quality control features for crowdsourcing experiments and fine-grained search logs\footnote{Behaviors logged include document dwell time, clicked documents, mouse hovers, document snippets shown on screen, bookmarked documents, etc.}. Figure~\ref{fig:interface} showcases the user interface we designed for our experiments. The eight main components are listed here. \circled{1}~ denotes the query box (without query auto-completion). \circled{2}~ represents the countdown timer to help our participants gauge the remaining minimum task time. \circled{3} highlights the task description. We show \circled{4} ten search results per page (each document can be saved \faBookmarkO{} to the \texttt{Saved documents} component for later usage, or hidden \faBan{} from future SERPs). Pagination is enabled \circled{5}. \circled{6}~shows the scaffolding component, with the \progress{} variant illustrated here (complete with yellow progress gradients). \circled{7}~shows the list of all issued queries so far in the search session, and \circled{8}~shows the list of all documents \textit{saved} so far in the search session. It should be noted that interface components \circled{6}, \circled{7}~and \circled{8}~provide scrollbars to scroll through content in each component. In the remainder of this section, we focus our discussion on our scaffolding variants, after introducing the approach behind our topical outlines.

\subsection{Topical Outlines}
A key ingredient of all our scaffolding strategies are the topical outlines for each learning topic (cf. Figure~\ref{fig:interface}, where the scaffolding component shows part of the outline for the topic \texttt{Ethics}). Effective outlines are typically hierarchical in nature~\cite{Hill2001education,glynn1977outline}, and follow a \emph{specific order} (ideally one that is best suited to master the topic). By providing such structure, we can point a learner toward a list of \emph{subtopics}---or topical aspects---that are important to the main topic.

\begin{figure}[t!]
\centering
  \includegraphics[width=0.6\linewidth]{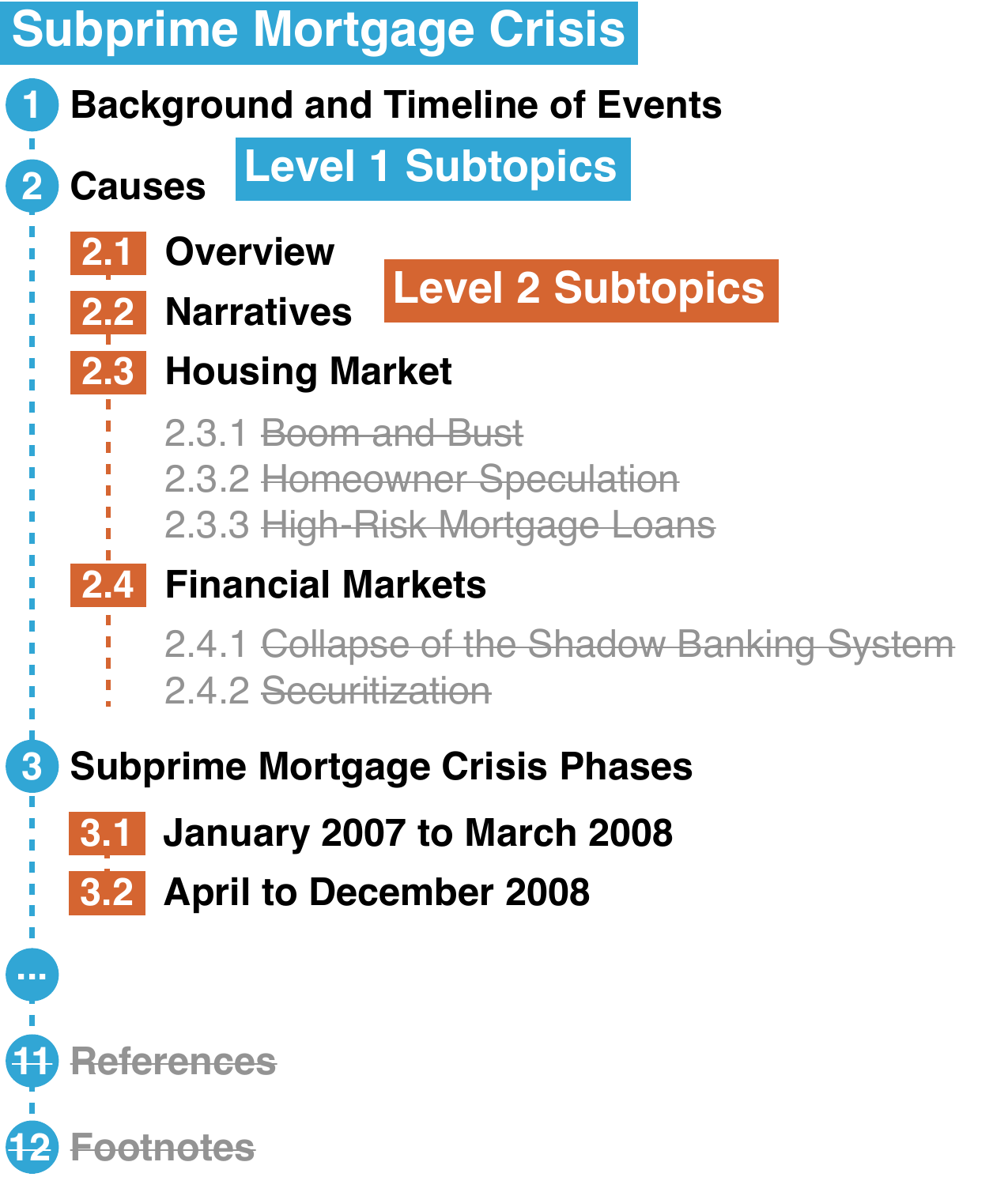}
  \caption{Hierarchical topic structure for the topic \textit{Subprime Mortgage Crisis}. Topic and structure derived from \car{} 2017~\cite{TRECCAR2017}. Note that third level subtopics (and deeper) and footnotes/references are excluded (illustrated in the figure by use of
      \vspace*{-4mm}
\sout{strikethroughs}).}
  \label{fig:topic structure}
\end{figure}

Such outlines can either be created by instructors~\cite{sharma2007scaffolding, belland2017instructional} or automatically (this is known as \emph{outline generation}~\cite{zhang2019outline}). The latter is desirable as it is scalable and not dependent on the availability of a domain expert---this is however a nontrivial challenge. For this reason, we rely upon manually created outlines for this study. More specifically, we used the heading structure of the corresponding Wikipedia article for each of our topics, as provided by the \car{} 2017 dataset~\cite{TRECCAR2017}\footnote{We note that topical outlines can also be extracted from text books or online courses; we picked Wikipedia here as source of our outlines since these are naturally hierarchical and readily available in the \car{} dataset (outlined in more detail in Section~\ref{sec:experiments}).}. This can be considered as employing a \emph{crowd of experts}~\cite{luyt2012inclusivity} for creating the outline. A concrete example outline from Wikipedia for the \emph{subprime mortgage crisis} topic is shown in Figure~\ref{fig:topic structure}. Each outline was manually cleaned; we only consider subtopics in our outline up to two levels deep (we refer to those levels as $L1$ and $L2$, cf. Figure~\ref{fig:topic structure}) and we remove generic subtopics that occur across a range of topics (such as \emph{References}).

\subsection{Variant \qe{}}
Scaffolding can be incorporated in different ways within a search system. It can be incorporated in the frontend (as we explore with \visual{} and \progress{}), or the backend. In the backend, we can either modify the retrieval function (as proposed by Syed and Collins-Thompson~\cite{syed2017retrieval, syed2018exploring}), or reformulate the \textit{to-be-submitted} queries. We chose the latter setup, as this is agnostic to the employed search engine (\textit{Bing} in the present study, via the \textit{Bing Search API}). More specifically, we reformulated each user query by appending the topic name (e.g. \emph{subprime mortgage crisis}) and one of the $L1$ subtopics (e.g. \emph{causes}) before submitting it to the search backend. Which subtopic we appended was dependent upon the \emph{time} the query was submitted during the search session. Each $L1$ subtopic was considered \emph{active} an equal amount of time. For example, for a search session estimated to last 30 minutes\footnote{As we set a minimum task time of 30 minutes in our study this is a reasonable setup.}, a topic with six $L1$ subtopics will have each subtopic active for five minutes. We chose to only include $L1$ topics here, as: \textit{(i)} the inclusion of $L2$ topics (of which there are usually two or three times as many) would lead to too many topical changes in a short period of time; and \textit{(ii)} the returned search results would often be overly specific. We kept the order of the subtopics as present in the topical outline intact. The search interface the study participants see in this variant is as shown in Figure~\ref{fig:interface}, but without the \circled{6} scaffolding component. Finally, we note that the \control{} variant has the same user interface as \qe{}, but no automatic query expansion is employed. Additionally, the participants had no visual indication that their queries were modified.

\subsection{Variant \visual{}}
As already mentioned, the next two scaffolding techniques are focused on changes to the frontend. Here, we explore to what extent making learners \emph{explicitly aware} of the topical outline impacts their search behaviors and knowledge gains. The first variant, \visual{}, is as seen in Figure~\ref{fig:interface}, though \emph{without} the yellow progress gradient (i.e. component \circled{6} is static, with solid blue backgrounds throughout). The scaffolding component contains the topic name (here: \texttt{Ethics}) and a list of $L1$ and $L2$ subtopics in order. As previously mentioned, the component has fixed dimensions, but can be scrolled at anytime. While the task description does not point explicitly to the component (as seen on the right of Figure~\ref{fig:interface}), we do introduce the component in an interactive tutorial before the start of the search session as follows:

\begin{quote}
\emph{This is a list of important subtopics. Each sub-topic can itself be broad enough to be split into several sub-topics. Explore the subtopics as much as you can.}
\end{quote}

The intuition behind this scaffolding choice is that learners that are pursuing a given list of curated subtopics should achieve higher knowledge gain than those searching without this guidance. 

\subsection{Variant \progress{}}
While \visual{} presents a static component to the learner which does not change during the search session, in \progress{} we provide feedback about the learners' progress throughout the search session. To do this, we estimate the exploration of each subtopic, and display this information as a progress bar as shown in Figure~\ref{fig:interface}, inspired by~\citet{umemoto2016scentbar}. In contrast to their approach, we cannot precompute the match of each document in the corpus to each subtopic (as we are using the open web, rather than a static corpus). The computation of how a list of viewed documents contributes to the progress of each subtopic is therefore nontrivial. This needs to happen in (near) real-time to avoid a noticeable lag.

Each time 10 search results (documents) are retrieved from the Bing Search API for a given query, we compute, for each document/subtopic pair the semantic similarity between them. To this end, we tokenize both document and subtopic\footnote{For tokenization we employ \url{https://github.com/huggingface/tokenizers}.}, and extract their sentence embedding using a pre-trained BERT-base model~\cite{devlin-etal-2019-bert}\footnote{Here, we follow the recommendations proposed by the authors of BERT of averaging all token embeddings from the second-to-last layer: \url{https://github.com/google-research/bert/issues/71}.}. We then  compute the cosine similarity between both embeddings, and that score, between 0 and 1, is used to increase the progress bar for the respective subtopic \emph{if the user views the respective document}. As this pairwise operation is expensive to do in near real-time (e.g. for the topic \emph{`noise induced hearing loss'} with 27 subtopics, we have to compute 270 document/subtopic similarities each time), we employ two additional filters that can be computed quickly: \textit{(i)} we remove documents with fewer than 50 tokens from consideration (there is little to learn in those cases), as well as \textit{(ii)} documents which contain less than 20\% of the unique terms in the section of the Wikipedia article for the subtopic. 

Thus, the similarity score of document $D_i$ for subtopic $t_j$ can be computed as follows:
 \begin{equation*}\label{eq:score}
    S(D_i,t_j) = 
    \begin{cases}
      \frac{\phi(D_i)\cdot \phi(t_j)}{||\phi(D_i)||\times||\phi(t_j)||}, & \text{if}\ |D_i| > 50 \wedge \frac{|D_i\cap t_j|}{|t_j|} > 0.2 \\
      0, & \text{otherwise}
    \end{cases}
  \end{equation*}
where $\phi(\cdot)$ is the embedding operation described before. Each viewed document can thus contribute to the progress score of multiple subtopics. We a consider subtopic's progress bar completely \textit{`filled up'} when the aggregate similarity score reaches $10$. This is a constant that is determined based on the search session length, and the number of subtopics present.

\section{User Study Setup}\label{sec:experiments}
Having outlined our scaffolding variants, we now consider the overall study setup, including a discussion on our choice of topics, the metrics we employ to measure learning gain, our study participants, and the workflow we followed.

\subsection{Topics}\label{sec:topics}
We used a subset of the 117 training topics from the \car{} 2017~\cite{TRECCAR2017} dataset. This dataset is a set of outlines, extracted from Wikipedia headings, with the original goal being to find relevant passages for each of these headings. This structure makes this dataset a good match for this task, since it already provides the required hierarchical topical outlines. 

We extracted the 100 topics whose topical outlines have at least two hierarchy levels, and then filtered those to an initial set of 48 by discarding topics that lack complexity. Of those, we picked 10 topics based on their difficulty and complexity, judged by 17 STEM graduate students\footnote{Each assessor received all 48 topics, in a randomized order, and was asked to select the 10 that appeared most difficult to them for learning about. Finally, the 10 topics selected most often were chosen as our topic set.}. Finally, we removed 3 topics: \emph{`Norepinephrine'}, as the Wikipedia page of the topic was mostly comprised of images; \emph{`research in lithium ion batteries'}, which contains a much larger number of subtopics (almost 50) than our other topics; and \emph{`theory of mind'}, which showed to be too easy, as almost no study participant was assigned to it (cf. Section~\ref{sec:Workflow} for how users were assigned to each topic). In the end, we worked with the remaining 7 topics, which are listed in Table~\ref{tab:studystats}. Each of the topics selected has between 11 and 27 subtopics. The choice for the most difficult topics was made so that we could maximize the potential learning of the participant during the experiment, and that any knowledge gained would be clearly apparent.

In order to measure users' learning gains, we followed the established approach of resorting to a pre- and a post-test of important concepts related to a topic~\cite{moraes2018contrasting,roy2020chiir,syed2017optimizing,gadiraju2018analyzing,yu2018predicting} (i.e. users are queried about their knowledge of the concepts \textit{before and after} the search session). In line with previous works, we resorted to a vocabulary knowledge test as the mean to evaluate domain knowledge. To this end, two of the authors manually selected 10 concepts per topic (listed in Table~\ref{tab:topicConcepts}) from the corresponding Wikipedia article---after an initial list of 100 candidate unigram/bigram concepts were automatically extracted using the highest IDF scores, computed on the \car{} 2017 corpus (a subset from Wikipedia), post stopword removal. When choosing the concepts, we aimed to pick the most representative terms for each topic by analyzing the respective Wikipedia article. Some unigrams and bigrams were further combined when needed for context (e.g. \texttt{inquiry commission} $\xrightarrow{}$ \texttt{financial crisis inquiry commission}) and stopwords were also re-introduced when needed (e.g. \texttt{overstimulation hair cells} $\xrightarrow{}$ \texttt{overstimulation \textbf{of} hair cells}).

\DeclareRobustCommand{\hleasy}[1]{{\sethlcolor{easy}\hl{#1}}}
\DeclareRobustCommand{\hldiff}[1]{{\sethlcolor{difficult}\hl{#1}}}
\begin{table}[!tb]
    \caption{Overview of the 10 concepts per topic utilized in the pre- and post tests. Highlighted are the easiest and most difficult two concepts per topic: marked in \hldiff{orange} (\hleasy{yellow}) are the two concepts of each topic with, on average, the \hldiff{lowest} (\hleasy{highest}) post-test knowledge scores.}
    \label{tab:topicConcepts}
    \centering
    \tiny
    \begin{tabular}{p{1.5cm}p{6cm}}
    \toprule
    
         Business cycle & 
        \hleasy{economic cycles}, \hldiff{distribution cycles}, swing cycle, \hleasy{wage cycle}, marxist model, endogenous causes, \hldiff{friedman}, capital profitability, model recession, austrian school \\

         \rowcolor{lightgray!22}
         Ethics &
         anarchist ethics, descriptive ethics, \hleasy{normative ethics}, relational ethics, virtue ethics, ethical resistance, \hleasy{consequentialism}, epicurean ethics, \hldiff{ethics feasible}, \hldiff{ethics spheres} \\

        Genetically modified organism &
      \hleasy{transgenic}, genomes, \hleasy{selective breeding}, microinjection enzyme, chromosome, plasmid, \hldiff{myxoma}, kanamycin, severe combined immunodeficiency, \hldiff{Leber's congenital amaurosis} \\
      
          \rowcolor{lightgray!22} 
      Irritable bowel syndrome &
    \hldiff{bifidobacteria infantis}, mesalazin, bile acid malabsorption, selective serotonin reuptake inhibitors, gut-brain axis, antidepressants, \hleasy{laxatives}, \hleasy{probiotics}, celiac disease, \hldiff{epithelial barrier} \\
    
    Noise-induced hearing loss &
    acoustic trauma, discomfort threshold, cochlear damage, audiogram, \hleasy{overstimulation of hair cells}, noise conditioning, \hldiff{excitotoxicity}, \hldiff{OSHA}, sensorineural hearing loss, \hleasy{tinnitus}, threshold shift \\
    
    \rowcolor{lightgray!22}
    Radiocarbon dating considerations &
    carbon exchange reservoir, isotopic fractionation, \hldiff{polarity excursion}, carbonate, \hldiff{geomagnetic reversals}, mass spectrometry, upwelling, \hleasy{radiocarbon}, \hleasy{neutrons}, photosynthesis pathways \\
    
    Subprime mortgage crisis  & 
    \hleasy{mortgage}, subprime, \hldiff{financial crisis inquiry commission}, securities, \hldiff{ben bernanke}, investment banks, \hleasy{housing bubble}, lehman brothers, foreclosures, default \\
         

    \bottomrule
    \end{tabular}
    \vspace*{-4mm}
\end{table}

\begin{table*}[!htb]
\caption{Overview of the topics used in our study, with associated statistics. Two-way ANOVA tests revealed no significant differences in average number of queries between topics ($F(6, 99) = 2.01, p=0.07$), or between the average number of bookmarks ($F(6, 99) = 0.41, p=0.87$).
}
\label{tab:studystats}
\footnotesize
\begin{tabular}{L{3.6cm}|D{1.35cm}|D{1.35cm}|D{1.35cm}|D{1.35cm}|D{1.35cm}|D{1.35cm}|D{1.35cm}}
\toprule
&
\shortstack{$^1$\textbf{Business}\\\textbf{cycle}} &
\shortstack{$^2$\textbf{Ethics}} &
\shortstack{$^3$\textbf{Genetically}\\\textbf{modified}\\\textbf{organisms}} &
\shortstack{$^4$\textbf{Irritable}\\\textbf{bowel}\\\textbf{syndrome}} &
\shortstack{$^4$\textbf{Noise}\\\textbf{induced}\\\textbf{hearing loss}} &
\hspace*{-1.5mm}\shortstack{$^6$\textbf{Radiocarbon}\\\textbf{dating}\\\textbf{considerations}} &
\shortstack{$^7$\textbf{Subprime}\\\textbf{mortgage}\\\textbf{crisis}} \cr
\midrule
\textbf{Level 1 subtopics} &  4  & 6  & 5 & 10 & 8 & 4 & 8  \\ 
\textbf{Level 2 subtopics} &  15  & 12  & 6 & 15 & 19 & 8 & 19 \\ \midrule
\textbf{Study participants}        &  16 & 20 & 15 & 15 & 19 & 21 & 20\\ 
\textbf{Participants for \control{}}  & 3 & 3 & 4 & 4 & 5 & 6 & 5 \\
\textbf{Participants for \qe{}} & 3 & 5 & 3 & 3 & 3 & 5 & 6\\
\textbf{Participants for \visual{}} & 4 & 5 & 4 & 3 & 4 & 6 & 7 \\
\textbf{Participants for \progress{}} & 6 & 7 & 4 & 5 & 7 & 5 & 2 \\ \midrule
\textbf{Average number of queries} &  $11.1 (\pm 6.4)$ & $11.4 (\pm 8.0)$ & $6.6 (\pm 3.9)$ & $10.1 (\pm 8.8)$ & $7.9 (\pm 6.6)$ & $7.8 (\pm 5.5)$ & $5.8 (\pm 3.4)$\\ 
\textbf{Median number of queries} &  9.5 & 9.5 & 6.0 & 7.0 & 7.0 & 6.5 & 5.0\\
\textbf{Average number of bookmarks} & $6.9 (\pm 6.4)$ & $8.8 (\pm 6.1)$ & $6.1 (\pm 3.5)$ & $7.7 (\pm 6.9)$ & $10.1 (\pm 11.0)$ & $10.0 (\pm 22.6)$ & $5.5 (\pm 5.9)$\\
\textbf{Median number of bookmarks} &  4.5 & 7.0 & 6.0 & 5.0 & 5.0 & 5.0 & 4.0\\
\bottomrule
\end{tabular}
\end{table*}

\subsection{Metrics}

We evaluate the knowledge gain of a concept by utilizing the \textit{Vocabulary Knowledge Scale (VKS)}~\cite{wesche1996assessing} across four levels (in line with~\cite{roy2020chiir,moraes2018contrasting}): 
\begin{enumerate}
    \item \textit{I don't remember having seen this term/phase before.}
    \item \textit{I have seen this term/phrase before, but I don't think I know what it means.}
    \item \textit{I have seen this term/phrase before and I think it means ...}
    \item \textit{I know this term/phrase. It means ... }
\end{enumerate}{}

This means that in both the pre- and post-tests, study participants were asked to rate themselves on their knowledge levels of each concept. Note that a self-assessment of \textit{(3)} or \textit{(4)} requires participants to write down a definition of the concept in their own words, which in turn allows us to grade the quality and reliability of the self assessment. It's also worth mentioning that the participants were not aware, at the start of the experiment, that the same questions would be asked again in the post-test, as this could influence their search behavior.

In order to compute the learning gain, we assign a score of \texttt{0} to both knowledge levels \textit{(1)} and \textit{(2)}. Since level \textit{(3)} indicates the participant is not certain about a concept's meaning, we assign it a score of \texttt{1}. Choosing level \textit{(4)} indicates the participant is confident in their assessment, and we assign it a score of \texttt{2}.\footnote{We note that this scoring scheme is equivalent to the \emph{fine-grained setup} employed by~\citet{moraes2018contrasting}.} 

\begin{figure}[!tb]
    \centering
    \includegraphics[width=0.95\linewidth]{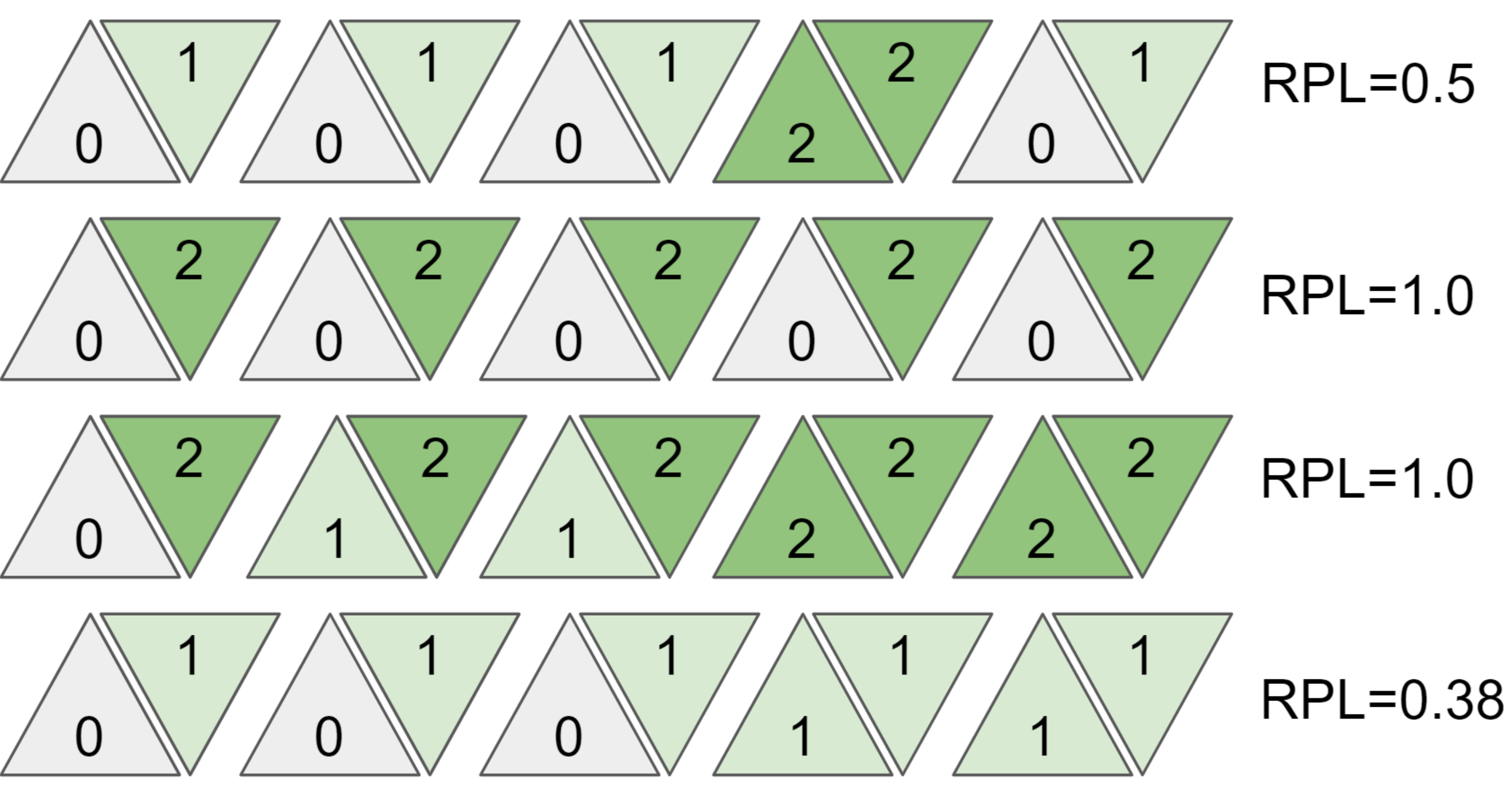}
    \caption{RPL examples: $\bigtriangleup$ represents $vks^{pre}$ and $\bigtriangledown$ represents  $vks^{post}$. Here, $n=5$.}
    \label{fig:rpl_example}
    \vspace*{-5pt}
\end{figure}

In line with~\cite{syed2018exploring,syed2017retrieval,moraes2018contrasting,colt2011measuring, shefelbine1990student}, we utilize \textit{Realized Potential Learning (RPL)} as our main learning gain metric. RPL depends on the \textit{Absolute Learning Gains (ALG)} which is measured in terms of either the number of new concepts learned (indicated by a score change of \texttt{0} to \texttt{1} or \texttt{0} to \texttt{2} from pre-test to post-test), \textit{or} the number of concepts they became more confident at (indicated by a score change of \texttt{1} to \texttt{2} from pre-test to post-test). RPL normalizes ALG by the \textit{maximum possible learning potential (MLG)}, which is \texttt{2} if the pre-test score is \texttt{0} or \texttt{1} if the pre-test score is \texttt{1}. And thus, for $n$ concepts: 

\begin{equation*}
\begin{aligned}
    ALG & = \frac{1}{n}\sum^n_{i=1}max(0, vks^{{post}}(v_i) - vks^{pre}(v_i))\\
        MLG &= \frac{1}{n}\sum^n_{i=1} 2 - vks^{pre}(v_i) \\
        RPL &= \frac{ALG}{MLG}.
\end{aligned}
\end{equation*}
Here, $vks^X(v_i)$ is our assigned score of concept $v_i$ (\texttt{0}, \texttt{1} or \texttt{2}), $X$ is either $pre$ or $post$ and $n=10$. 
Intuitively, RPL measures the percentage of knowledge gained from the total possible knowledge to be gained. To provide the reader with some intuition, we provide concrete examples of how pre/post-test scores translate into RPL in Figure~\ref{fig:rpl_example}.

We note that $ALG$ and $RPL$ are not the only possible metrics to measure learning. Instead of treating each concept in the same manner, \textit{difficulty weighted learning gains} can be computed too (as for instance done by~\citet{syed2018exploring}, where vocabulary items such as \texttt{earth} and \texttt{temperature} were mixed with more technical vocabulary items). Based on the manner we selected our concepts, we do not believe this to be necessary as they are similarly difficult. Some prior works have also manually annotated participants' summaries or mind maps to derive qualitative and quantitative metrics~\cite{o2020role,wilson2013comparison,liu2019investigation}. We leave the analyses of the user summaries we collected in this manner for future work.

\subsection{Study Workflow}\label{sec:Workflow}
\begin{figure}[!tb]
    \centering
    \includegraphics[width=0.95\linewidth]{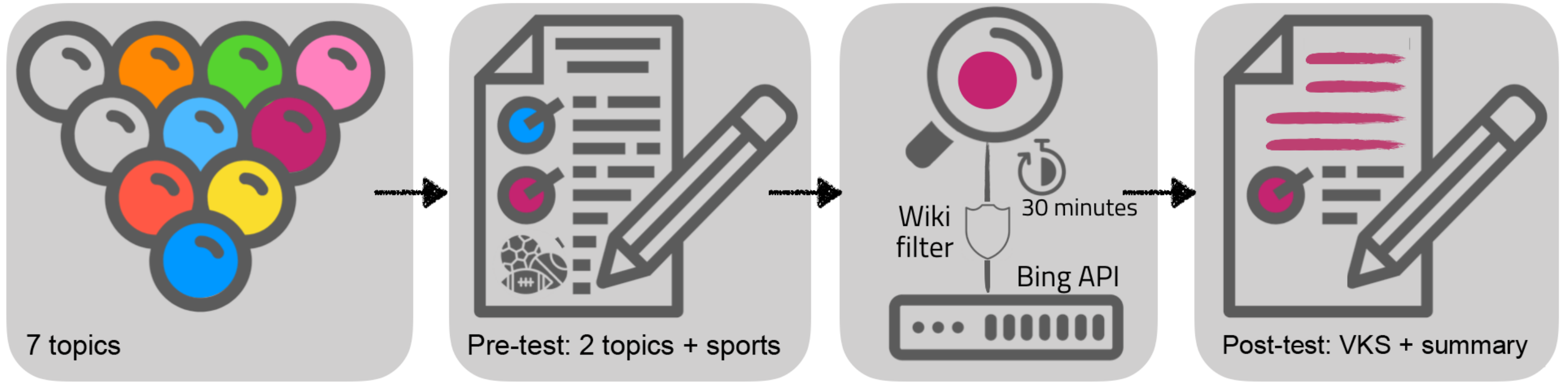}
    \caption{Overview of the flow of our user study.}
    \label{fig:study}
\end{figure}

The flow of our user study is presented in Figure~\ref{fig:study}; it is implemented within our \searchx{} instance. When a participant enters the study, two of our seven topics are randomly selected. In addition to this (and to weed out non-complying crowd workers), we add the topic \emph{`sports'} to the pre-test as we expect reasonable participants to demonstrate high knowledge levels on this topic. The pre-test thus consists of 30 VKS questions in total. We rejected crowd workers that score lower on \emph{`sports'} than the other two topics. The topic they know the least about is then chosen as the one to learn about during the search session. We introduced the simulated learning task as shown in Figure~\ref{fig:interface}, item \circled{3}. The minimum task time was set to thirty minutes. We also filtered any web document returned from the Bing Search API that either came from a Wikipedia domain, or domains that are known clones of Wikipedia\footnote{We blacklisted a total of 72 domains. All subtopics were submitted to the Bing Search API, with the top 10 results returned. Each result was then inspected to determine whether it came from a Wikipedia clone in our blacklist.}. Wikipedia and its clones were excluded as we drew our topical outlines from the relevant Wikipedia article -- the said Wikipedia article would therefore be the best to read. While for a large portion of topics Wikipedia is a great tool for learning, we cannot expect good Wikipedia pages for all topics, especially niche or highly specific topics. Therefore, we believe that the formulation outlined in this section is still a reasonable search task.

During the search session, participants could search, view and bookmark documents. In the post-test, we asked them again about their knowledge on the ten concepts for their topic. In addition, we asked them to write a summary (100 words minimum)\footnote{Specifically, we phrased this as: \textit{``Your professor also asks you to write a summary of what you learned about the topic you searched about. This summary should be enough that your fellow students that read it can get a first idea of what the topic is, without having to search for it themselves. Please write your summary here (at least 100 words).''}} about the topic. We note here that the knowledge tests require understanding, but no application or synthesis (i.e. higher-level cognitive processes of learning~\cite{Anderson2001}) of the materials---here we make the bookmarked documents available to our participants. 

\subsection{Study Participants}

We conducted our study on the \textit{Prolific Academic platform}\footnote{\url{https://www.prolific.co/}} across three days. In order to ensure responses of high quality, we required our participants to have at least 15 previous submissions, an approval rate of 90+\% and be native English speakers. The study took about an hour to complete and participants were reimbursed with \textsterling 6 per hour. 144 participants completed our study. We had to reject 18 participants (leading to $N=126$ valid participants) as they had completed more than three browser tab changes (we enforced this rule to ensure our participants actively using our search system instead of running down the timer). Of the valid participants, 65 were male, 59 female (2 withheld gender information) with a median age of 27 (minimum 18, maximum 63). 44 participants reported as highest formal education level a high school degree, 47 reported a Bachelor's degree and 20 had a Master's degree. The remaining 15 participants indicated other educational levels.

In Table~\ref{tab:studystats}, we report the number of participants per topic. The maximum number of participants were assigned to the topic \emph{`radiocarbon dating considerations'} (21), while the minimum were assigned to \emph{`genetically modified organisms'}  and \emph{`irritable bowel syndrome'} (15). 
The table also contains statistics on the number of queries and bookmarks per topic, indicating that our study participants actively engaged in the search session. The median number of queries ranges from 5 to 9.5, with the median number of bookmarks ranging from 4 to 7 respectively across the topics.

At the end of the data collection, we had collected answers for 1260 VKS questions and 126 essays. In order to determine the quality of the VKS self-assessments, we sampled 100 concept definitions written by our participants: 50 for knowledge levels \textit{(3)} and \textit{(4)} respectively. Two annotators labeled them as \textit{correct}, \textit{partially correct}\footnote{Partially correct definition example of \emph{tinnitus} (i.e., noise induced hearing loss topic): \textit{``hearing loud sounds in one's ears.''}} and \textit{incorrect}\footnote{Incorrect definition example of \emph{genomes} (genetically modified organism topic): \textit{``the amount of chromosomes.''}}. We find that 25.2\% of the vocabulary scores self assessed as \textit{(3)} were correct; 65.9\% were partially correct; and the remaining 8.9\% were incorrect. Among the definitions self assessed as \textit{(4)}, 64.8\% were correct, 28.9\% were partially correct and the remaining 6.3\% were incorrect. Based on these numbers, we consider the self-assessment to be largely reliable. Thus, we report RPL based on self-assessed vocabulary knowledge levels.
\section{Results}
We now turn to addressing our research questions. In terms of statistical tests reported within this section, we performed two-way ANOVA tests (with two factors: the intervention type and topic), followed by a post-hoc two-way Tukey HSD pairwise test in case of significance ($\mathbf{p<0.05}$).\footnote{For further investigations, An anonymized version of the data is available at \url{https://github.com/ArthurCamara/searchx-scaffolding}}

\subsection{RQ1: Impact of Scaffolding on Learning}
\begin{figure}[!tb]
    \centering
    \includegraphics[width=\linewidth]{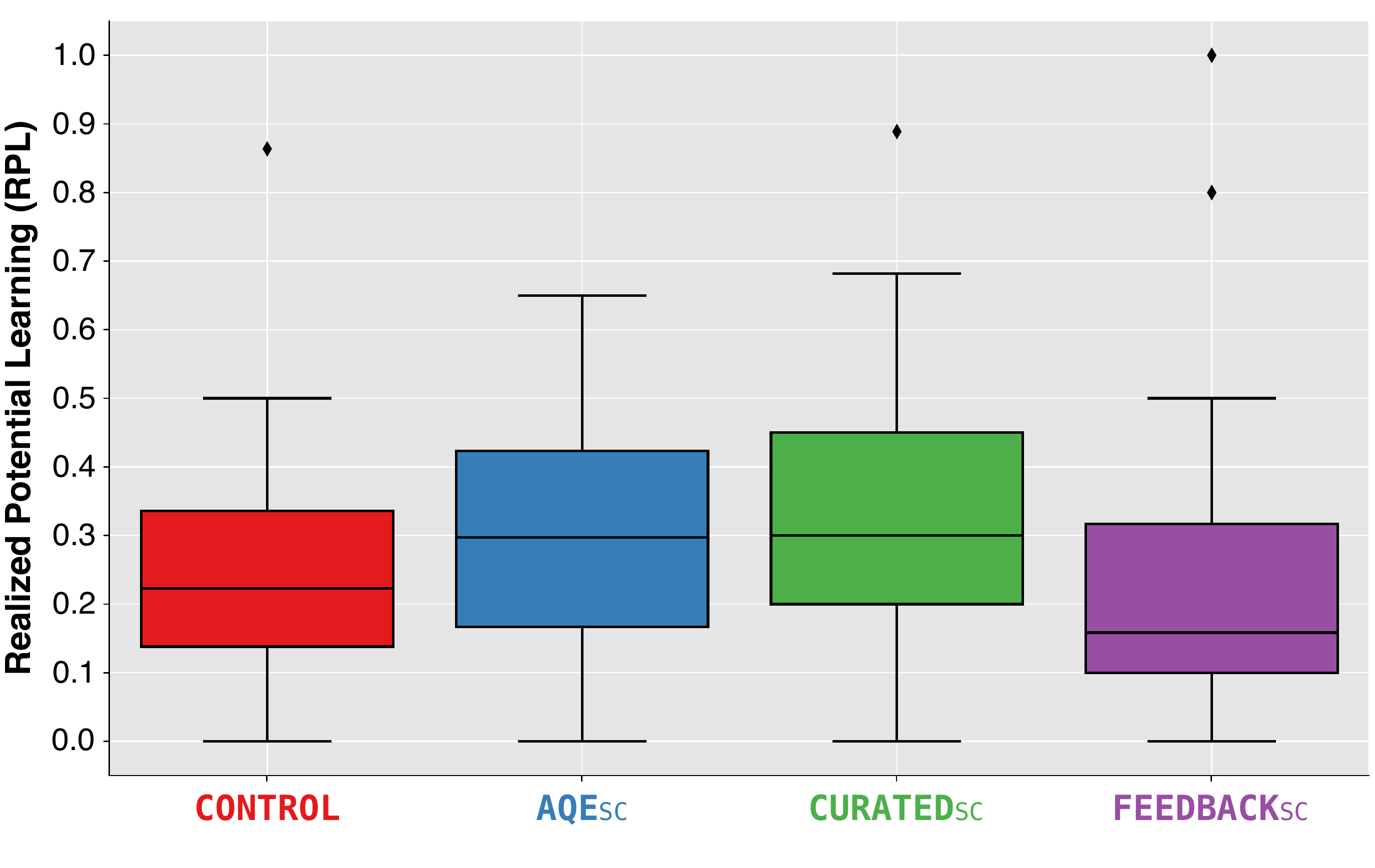}
    \vspace*{-4mm}
    \caption{RPL over the four different conditions.}
    \vspace*{-4mm}
    \label{fig:RPL}
\end{figure}

\begin{figure}[!tb]
    \centering
    \includegraphics[width=0.9\linewidth]{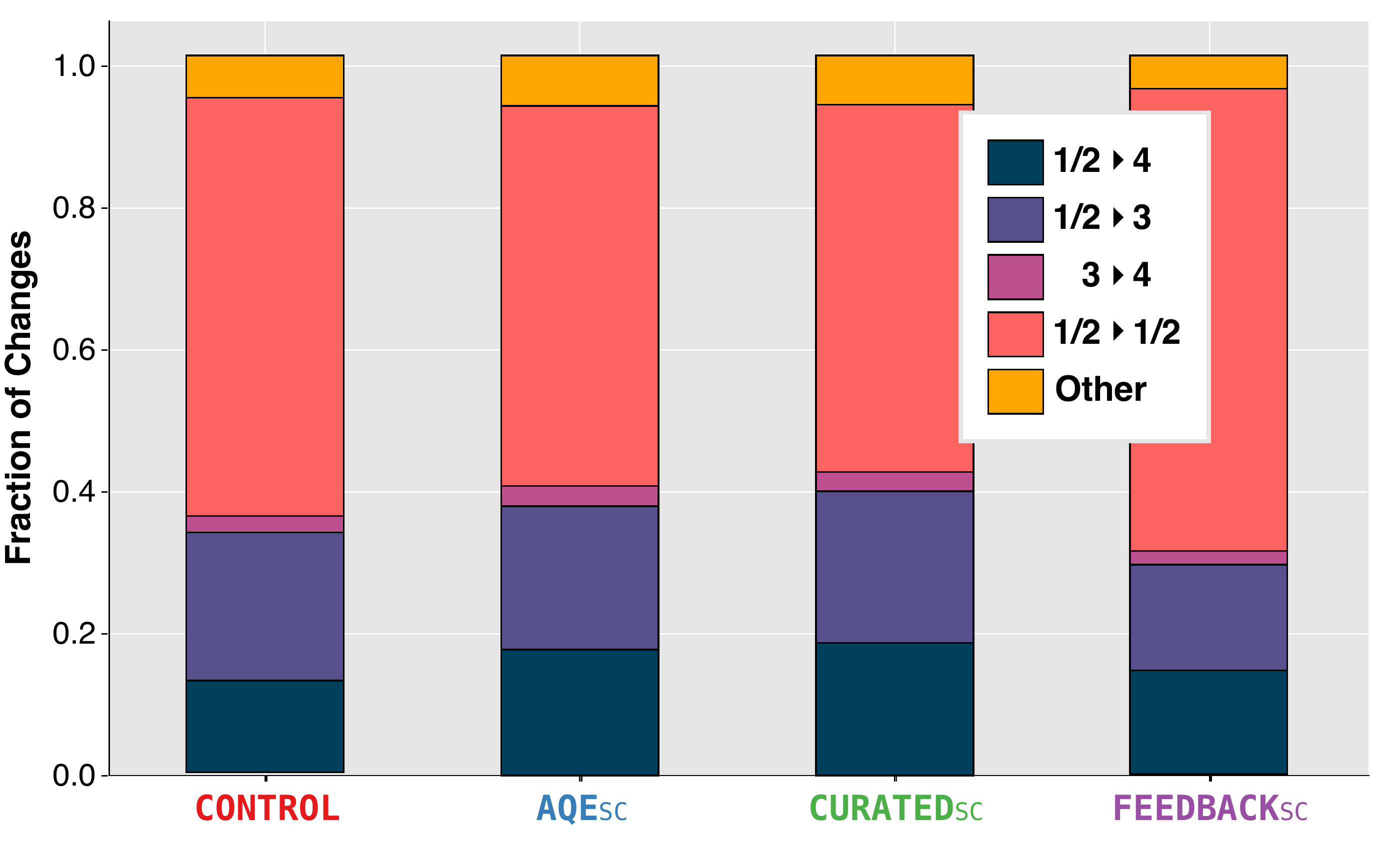}
    \vspace*{-4mm}
    \caption{Fraction of change in vocabulary tests.}
    \vspace*{-5mm}
    \label{fig:jump}
\end{figure}

In Figures~\ref{fig:RPL} and~\ref{fig:jump}, we present the RPL across the four conditions (each one with between 28 and 36 participants, and an average search session duration\footnote{We compute the search session duration as the time between the first query issued and the last viewed document closing.} of more than 36 minutes, cf. Table~\ref{tab:summarystats}), and a more fine-grained presentation of the knowledge changes.

\begin{table*}
\caption{Mean ($\pm$ standard deviations) of RPL and search behavior metrics across all participants in each condition. $^\dagger$ indicates two-way Anova significance, while $^\mathcal{C}, ^\mathcal{A}, ^\mathcal{U}, ^\mathcal{F}$  indicate post-hoc significance (TukeyHSD pairwise test, $\mathbf{p < 0.05)}$ increases vs. \control{},  \qe{},  \visual{} and \progress{} respectively.}
\label{tab:summarystats}
\footnotesize
\begin{tabular}{p{0.15cm}L{7.5cm}|cccc}
\toprule
& &
\control{} & \qe{} & \visual{} & \progress{} \\
\midrule
I. & \textbf{Number of participants} & 30 & 28 & 33 & 36 \\
II. & \textbf{Search session duration (minutes)} & $36m33s (\pm 12m15s)$ & $39m59s (\pm 10m59s)$ & $41m31s (\pm 13m6s)$ & $38m15s (\pm 12m46s)$
 \\
\midrule
III. & \textbf{RPL} & $0.26 (\pm 0.18)$ & $0.30 (\pm 0.16)$ & $0.31 (\pm 0.20)$ & $0.24 (\pm 0.24)$  \\

\midrule
IV. & \textbf{Number of queries$^\dagger$}        &  $5.13 (\pm 2.61)^{\mathcal{U}\mathcal{F}}$ & $5.29 (\pm 2.98)^{\mathcal{U}\mathcal{F}}$ & $11.09 (\pm 6.99)^{\mathcal{C}\mathcal{A}}$ & $11.86 (\pm 7.60)^{\mathcal{C}\mathcal{A}}$ \\ 

V. & \textbf{Fraction of query terms coming from topical outline$^\dagger$} &   $0.26 (\pm 0.28)^{\mathcal{U}\mathcal{F}}$ & $0.33 (\pm 0.31)^{\mathcal{U}\mathcal{F}}$ & $0.58 (\pm 0.34)^{\mathcal{C}\mathcal{A}}$ & $0.58 (\pm 0.29)^{\mathcal{C}\mathcal{A}}$  \\

VI. & \textbf{Fraction of topical outline terms used for querying$^\dagger$} &   $0.04 (\pm 0.04)^{\mathcal{U}\mathcal{F}}$ & $0.05 (\pm 0.04)^{\mathcal{U}\mathcal{F}}$ & $0.32 (\pm 0.23)^{\mathcal{C}\mathcal{A}}$ & $0.34 (\pm 0.24)^{\mathcal{C}\mathcal{A}}$  \\

VII. & \textbf{Average time between queries (minutes)} &  $5m57s (\pm 5m26s)$ & $6m31s (\pm 8m31s)$ & $3m31s (\pm 2m45s)$ & $3m52s (\pm 4m40s)$ \\
\midrule
VIII. & \textbf{Average time between document close and next document load (secs.)} &  $60.15 (\pm 27.17)$ & $68.06 (\pm 33.44)$ & $74.42 (\pm 45.14)$ & $57.32 (\pm 39.13)$  \\ 

IX. & \textbf{Average document dwell time (secs.)} & $76.77 (\pm 51.14)$ & $100.61 (\pm 61.59)$ & $92.15 (\pm 97.60)$ & $55.33 (\pm 51.04)$  \\

X. & \textbf{Number of unique documents viewed$^\dagger$} & $14.77 (\pm 8.85)$ & $10.96 (\pm 4.08)^\mathcal{F}$ & $14.09 (\pm 7.95)$ & $18.50 (\pm 9.56)^\mathcal{A}$ \\

XI. & \textbf{Number of unique document snippets viewed$^\dagger$}&   $97.47 (\pm 47.37)^\mathcal{F}$ & $81.07 (\pm 44.58)^{\mathcal{U}\mathcal{F}}$ & $136.42 (\pm 76.97)^\mathcal{A}$ & $152.44 (\pm 84.23)^{\mathcal{C}\mathcal{A}}$  \\




\bottomrule
\end{tabular}
\vspace*{-3mm}

\end{table*}

Recall that RPL provides us insights into the amount of learning that has taken place with respect to the maximum possible amount of learning (which may differ per participant; some participants may have no prior knowledge of any of the ten concepts, while others have a good understanding of 2-3 concepts already). For the \control{} condition, the mean RPL is 0.26. Participants in both \qe{} and \visual{} on average report higher learning gains with an RPL of 0.3 and 0.31 respectively. To evaluate the impact of \qe{} in the set of retrieved documents, we collected the SERPs of both the original user-formulated and automatically reformulated queries, and found that, among the top 10 retrieved documents, an overlap on average of $1.5$ documents. That indicates that \qe{} had a great impact on how the SERP was presented.

Participants in the \progress{} condition had the lowest average RPL (0.24) as well as the highest standard deviation. This finding seems counter-intuitive, as the extra feedback available was hypothesized to be beneficial to the learning experience (as also envisioned, among others, by \citet{von2019metacognitive}). We provide a further investigation of possible reasons for this finding in Section~\ref{sec:analyses}. 


\begin{figure}[!tb]
    \centering
    \includegraphics[width=1\linewidth]{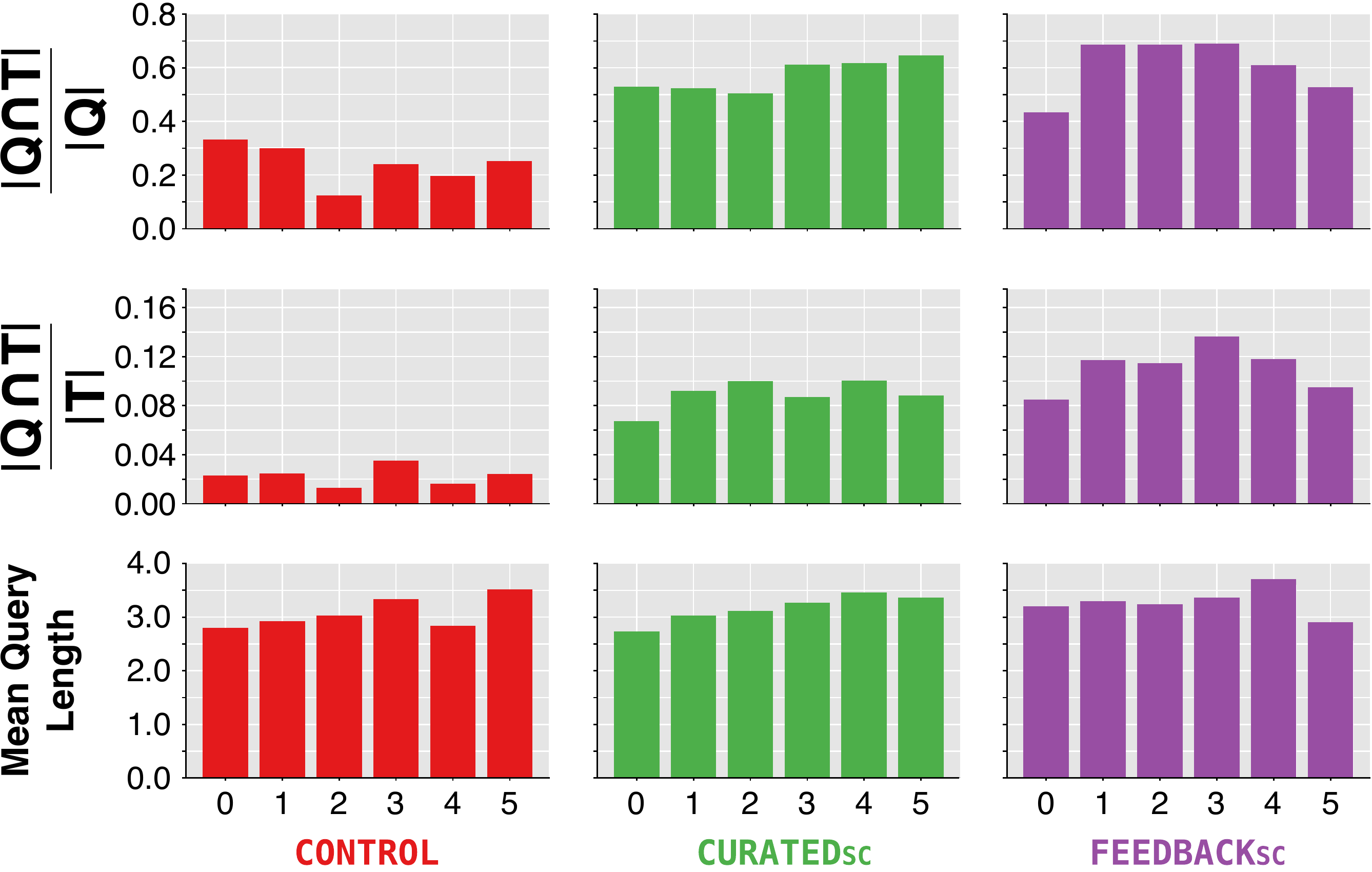}
    \caption{For each row: \textit{(top)} the fraction of query terms taken from topic outlines; \textit{(middle)} the fraction of topic outline terms used for querying; and \textit{(bottom)} the mean query length, over 5 minute blocks ($x$ axes) of the 30 minute search session, considering: \control{} \textit{(left)}; \visual{} \textit{(center)}; and \progress{} \textit{(right)}. Here, we consider the first query instance as the start of the first interval.}
    \label{fig:bigplots}
    \vspace*{-6mm}
\end{figure}

To further analyze how the conditions differ, we provide a detailed breakdown of the knowledge state transitions in Figure~\ref{fig:jump}. We are particularly interested in the transitions from states $1/2$ (where very little is known about a concept), to state $4$ (where the concept is completely understood). The percentage of concepts for which this holds is largest among \visual{} participants; similarly, the lack of knowledge increase (i.e. the transition $1/2 \rightarrow 1/2$) is smallest for this cohort. This result implies that the \visual{} cohort, on average, was most confident in their knowledge increase.

Overall, we conclude that there is a lack of evidence to support the conclusion that scaffolding increases participants' learning gains, despite the positive trends we observe for \qe{} and \visual{}. We found no significant difference ($F(3, 99) = 0.75, p=0.522$) between the four scaffolding conditions, which means that we cannot reject the null hypothesis that there is no learning gain difference among them. It is thus not as simple as introducing an outline or providing instantaneous feedback to yield reliable and large learning gains across a range of participants and across a range of topics.


\subsection{RQ2: Search Behavior Analyses}\label{sec:analyses}

Besides learning gains, we are also interested in the search behaviors of our participants. To answer our second research question, \textit{When scaffolding is introduced, to what extent does learners' search behavior change?}, we report a number of search behavior metrics (mean and standard deviations) in Table~\ref{tab:summarystats}. 

\subsubsection{Influence of visual scaffolds on querying.} 
Our participants in the \visual{} and \progress{} conditions issued significantly more queries (on average more than twice as many) than participants in the \control{} and \qe{} conditions (in line with~\cite{umemoto2016scentbar}). As a consequence, the average time between queries in those two conditions is much lower (less than four minutes on average, vs. more than six minutes on average) than in \control{} and \qe{}. We hypothesize that the readily available cues of what to query for enticed our participants to issue more queries, as they are aware of the various topical aspects. 
To validate this hypothesis---and in order to explore to what extent the participants in \visual{} and \progress{} made use of these visual cues---we determined: \textit{(i)} the percentage of unique query terms drawn from the topical outline; and \textit{(ii)}, the percentage of unique terms in the topical outline present in at least one submitted query. To this end, we converted the queries ($\mathcal{Q}$) and topical outlines ($\mathcal{T}$) into bags-of-words with normalization (stopword removal, capitalization, etc.), and computed $\frac{| \mathcal{Q} \cap \mathcal{T}|}{|Q|}$ as well as $\frac{| \mathcal{Q} \cap \mathcal{T}|}{|T|}$. The results in Table~\ref{tab:summarystats} (rows V \& VI) show clearly that the presence of the outline influences the querying behavior significantly: more than half the query terms are \textit{`borrowed'} from the topical outline in \visual{} and \progress{}, while this is the case for 33\% and 26\% on average for \qe{} and \control{} respectively where participants had no access to the outline. 
In addition, when considering the coverage of the topical outline by query term, we see once again that a much larger percentage of outline terms were queried at least once ($>30\%$ on average for \visual{} and \progress{} vs. $\leq5\%$ on average for the other two conditions) by participants in the variants with access to the outline. In the top two rows of plots in Figure~\ref{fig:bigplots}, we break down this comparison of query terms and topical outline terms further by splitting our search sessions into five minute intervals, and computing $\frac{| \mathcal{Q} \cap \mathcal{T}|}{|Q|}$ and $\frac{| \mathcal{Q} \cap \mathcal{T}|}{|T|}$ separately for each interval. We find that participants in the \control{} condition were not picking up more topical outline terms over time (despite the fact that they have read more documents on the topic by each passing interval). However, we do see a slight increase over time for \visual{} and \progress{}, which then drops again in the later stages of the search session.

\subsubsection{Too much feedback considered harmful.}

Previous works~\cite{roy2020chiir, moraes2018contrasting, gadiraju2018analyzing, yu2018predicting} have shown the number of queries issued to be a good proxy for learning. In our work, this finding holds for \visual{}, though not for \progress{}: on average, a similarly high numbers of queries were submitted, but the learning gains for \progress{} are low. For completeness, the bottom row of plots in Figure~\ref{fig:bigplots} shows mean query lengths across time: as the recorded search sessions progressed, queries tended to become longer.

We hypothesize that \progress{}, with its additional feedback to the participants, is counterproductive to their learning efforts due to \textit{the effects of gamification.} That is to say, instead of focusing on learning, participants are focused on trying to \textit{`fill up'} the progress bar. This leads to less self-reflection whilst reading documents as participants' focus is now on the progress of the scaffolding bar. Consequently, this causes a decrease in the learning gain. 
 
To empirically evaluate this hypothesis, we can look at the average document dwell time (Table~\ref{tab:summarystats}, row IX): it is on average 55 seconds in the \progress{} variant, which is significantly lower than that of the \visual{} and \qe{} variants (with an average document dwell time of 92 seconds and 100 seconds, respectively). At the same time, \progress{} participants viewed on average the most documents, and the most document snippets (Table~\ref{tab:summarystats}, rows X and XI). 
In addition, Figure~\ref{fig:bigplots} (middle row) shows that, as the search session progresses, participants from \progress{} tend to use more terms from the outlines than their \visual{} counterparts. 


To explain the large gap between the results of \visual{} and \progress{},~\citet{swinnen1990information} in a psychology study showed that learners who are presented with frequent feedback on their learning progress tend to learn less than others that do not. It is hypothesized that this is because this frequent feedback may impair their ability to \emph{reflect} on what they have learned. Similarly,~\citet{mayer2008increased} corroborate these findings in the setting of multimedia learning, demonstrating that too much extra information can distract learners from their core learning material. We believe that a similar effect may be in play here.


\section{Conclusions}
In this work, we have explored three strategies to introduce instructional scaffolding into a web search system with the goal of improving a learner's knowledge gain during the search process. These strategies were: \textit{(i)} automatic query rewriting (\qe{}) which is agnostic to the search backend; \textit{(ii)} a curated static topical outline (\visual{}); and \textit{(iii)} a curated topical outline with instant feedback on the exploration of the topic space (\progress{}). 

We conducted a user study with 126 participants and aimed to answer the following research questions:
\begin{enumerate}
\item[\textbf{RQ1}] Is scaffolding effective to increase learning outcomes?
\item[\textbf{RQ2}] How does the introduction of scaffolding change behaviors?
\end{enumerate}

Answering, \textbf{RQ1}, we do not find sufficient evidence to corroborate that any of the proposed methods significantly impacts learning outcome. However, we open a new research venue, showing that scaffolding significantly changes user behavior on a number of metrics. This is shown by our analysis answering \textbf{RQ2}, where we show that explicit scaffolding (namely \visual{} and \progress{}) significantly alters users behavior in a number of important search metrics, like dwell time, number of queries issued and number of clicks. This is important, and should lead to further investigation on how we can use this behavior difference to better support learners.

Additionally, we have speculated that the discrepancy in behavior  between \visual{} and \progress{}, albeit not significant, may be due to a gamification effect: instead of focusing on the task at hand (learning), participants are more focused on making progress on filling up their progress bars, and in the process lose sight of their goal. This is corroborated by the difference in dwell time, as the \progress{} condition led participants to skim the documents more than in other conditions (i.e. that condition had the lowest document dwell time) while spending more time on the SERP (highest number of document snippets viewed). Finally, we found that participants in the two conditions receiving the topical outline submitted more queries with many more query terms matching the terms in the topical outline.


From these results, there are several lines of future work to follow. Firstly, a better scaffolding component is needed: what type of interface/feedback to learners respond to best? In order to make this approach deployable in practice, we need to be able to \emph{automatically generate} hierarchical outlines for any learning-oriented information need instead of relying on manually curated outlines. Those outlines should preferably be personalized, depending on users' domain expertise and other user characteristics. While exploration into (non-personalized) automatic outline generation~\cite{zhang2019outline} is relatively new, it remains unclear whether such slightly noisy outlines are beneficial for users' learning outcomes. In addition, it remains to be seen to what extent the changes in user behavior hold across time (as for instance explored by~\citet{syed2018exploring}), and whether users remain engaged over time when a scaffolding component is permanently introduced on the search interface. Finally, we need to consider that we measured learning with a vocabulary knowledge task, which covers only the lowest cognitive levels of learning~\cite{anderson1993computer}. Is scaffolding beneficial for learners that face learning tasks that target higher cognitive levels of learning~\cite{kalyani2019understanding}?

\balance
\bibliographystyle{plainnat}
\bibliography{references}

\end{document}